\documentclass[12pt]{article}
 \usepackage[margin=1in]{geometry}
\usepackage[utf8]{inputenc}
\usepackage[T1]{fontenc}
\usepackage{graphicx}
\usepackage{rotating}
\usepackage{tablefootnote}
\usepackage{float}
\usepackage{caption}
\usepackage{setspace}
\usepackage{pdflscape}
\usepackage{textcomp}
\usepackage{url}

\usepackage[numbers,sort&compress]{natbib}
\usepackage{tabularx}
\usepackage{afterpage}
\usepackage{fancyhdr}
\usepackage[table]{xcolor}
\usepackage{array}
\usepackage{booktabs}
\usepackage{amsmath}
\usepackage{amssymb}
\usepackage{bm}
\usepackage{siunitx}
\usepackage{longtable}
\usepackage{pgfplots}
\pgfplotsset{compat=1.18}

\usepackage[hidelinks]{hyperref}

\graphicspath{{media/}{./}}
\renewcommand{\arraystretch}{1.12}
\definecolor{lightgray}{gray}{0.95}

  \title{The Economics of AI Training Data: A Research Agenda\thanks{Companion
  dashboard at \url{https://huggingface.co/spaces/midah/odl-training-data}.}}
  \author{
    Hamidah Oderinwale and Anna Kazlauskas\\                                    
    Open Data Labs\\                      
    \texttt{\{hamidah, anna\}@opendatalabs.xyz}
  }                                                                             
  \date{August 2026}
                          
\begin{document}
\maketitle

\begin{abstract}
Despite data's central role in AI production, it remains the least understood input. As AI labs exhaust public data and turn to proprietary sources through deals reaching hundreds of millions of dollars, research across disciplines has remained fragmented, resulting in no unified lens for studying the data supply chain \cite{Hopkins_2025}. We establish data economics as a coherent field through three contributions. First, we characterize data's distinctive properties (nonrivalry, context-dependence, and emergent rivalry through contamination) and trace historical precedents for market formation in commodities like oil and grain. Second, we present systematic documentation of AI training data deals from 2020 to 2025, revealing persistent market fragmentation, five distinct pricing mechanisms (from per-unit licensing to commissioning), and that most deals exclude original creators from compensation. Third, we propose a formal hierarchy of exchangeable data units (token, record, dataset, corpus, stream) and argue for data’s explicit representation in production functions. Building on these foundations, we outline four open research problems foundational to data economics: measuring context-dependent value, balancing governance with privacy, estimating data's contribution to production, and designing mechanisms for heterogeneous, compositional goods. 
\end{abstract}


\section{Introduction}

Data remains the least understood of the three inputs to the still vaguely defined AI production function, even as scaling laws \cite{hoffmann2022training} highlight its role in driving frontier capabilities alongside compute and algorithms. Most economic research on AI has emphasized macro outcomes such as labor and productivity effects, while neglecting the production side \cite{Huang2021-or}.\footnote{See also Reuel et al.'s \emph{Open Problems in Technical AI Governance} \cite{reuel2025openproblemstechnicalai} for related governance-level questions, though their focus is institutional rather than economic.}

In every technological revolution, understanding production-side economics has been prerequisite for understanding economic impact. During industrialization, understanding labor organization and supply chains was essential. During electrification, understanding energy infrastructure mattered. For AI, the same principle applies: we cannot understand its economic impact without understanding how it is produced. Yet current AI economics research focuses primarily on downstream effects, while treating the production function as a black box.\footnote{See Anthropic's Economic Index and Economic Futures Project \cite{anthropic2024economic,anthropic2024futures}, Stripe's Economics of AI Fellowship \cite{stripe2024fellowship}, and OpenAI's GDPval benchmark \cite{openai2024gdpval} as representative examples.}

This paper does not attempt to fully characterize AI's production function or resolve how data should be valued and allocated. Instead, it lays the groundwork for a formal field of data economics by: (1) documenting how data is currently exchanged and priced, (2) developing preliminary frameworks for representing data as a distinct factor of production, and (3) identifying open problems whose resolution would advance data economics as a field.

While foundation models have ingested much of the public internet \cite{villalobos2024data}, they use less than 0.01\% of the world's available data \cite{openminedAIdata}—roughly 99\% remains as ``dark data'' \cite{wikipediaDarkData} in proprietary databases, behind login walls, and in domain-specific corpora. As labs exhaust public data and turn toward proprietary sources—with deals reaching tens to hundreds of millions of dollars (see Table~\ref{tab:data-deals})—formal frameworks for understanding data's economic role remain underdeveloped despite growing activity across AI research, economics, law, and policy. These efforts rarely speak to one another.

This paper consolidates disparate insights into a coherent research agenda. Section 2 establishes why data resists standard economic treatment. Section 3 proposes a hierarchy of exchangeable units. Section 4 documents current pricing mechanisms. Section 5 sketches how data enters production functions. Section 6 identifies foundational research problems for data economics as an emerging field.

\section{Why data resists standardization}
\label{sec:properties}

Before examining how data is currently exchanged or how it might be integrated into production models, we must understand why it resists the market mechanisms that work for traditional factors of production. This section establishes three foundations: data's unique economic properties, historical precedents showing how heterogeneous assets became standardized, and the current state of research attempting to bridge these gaps.

\subsection{Economic properties and barriers to standardization}

Data is quite different from traditional commodities. It is nonrivalrous in principle: its reuse does not diminish its supply, and only partially excludable, since access can be restricted but copies are easily made. However, contamination effects (dataset aging, adversarial poisoning, benchmark leakage, preference leakage) and overuse create practical rivalry, reducing future utility \cite{zhang2024carefulexaminationlargelanguage,choi2025contamination,li2025preferenceleakage,huang2023poisoningsurvey}. While nonrival in principle, data has become effectively excludable as consent protocols restrict crawling and AI use \cite{longpre2024consent}.

Two barriers prevent data from following the standardization path that other heterogeneous assets have taken. First, the \emph{verification paradox}: quality and suitability cannot be assessed without examining data, yet examination enables copying, creating severe adverse selection where sellers cannot credibly signal quality and buyers cannot distinguish high-quality from low-quality data without access \cite{akerlof1970market}. The problem is particularly acute for data because, unlike physical goods, inspection grants perfect replication rather than mere knowledge. Second, \emph{legal opacity}: data's legal status (licensing rights, copyright clearance, consent validity) cannot be verified through inspection alone and remains uncertain even after investigation, as evidenced by ongoing disputes over data use and the difficulty in scaling standardized licensing agreements \cite{ropesgray2025unfairuse}. Together, the verification paradox and legal opacity create substantial obstacles to standardization---intermediaries and brokers become essential gatekeepers, raising transaction costs and fragmenting markets.

Beyond these verification barriers, data's value is highly context-dependent: it varies by buyer holdings, application, and combination with other datasets, and its highly differentiated nature prevents uniform pricing even when quality and legality are established \cite{cohen2016feature}.

\subsection{Historical precedents: How heterogeneous assets became standardized}

Data's heterogeneity is often cited as a barrier to treating it as a tradable asset. Yet history shows that even highly heterogeneous resources can become measurable and tradable through the development of standards, exchanges, and verification mechanisms when economic pressure demands it.

Consider the corporate form itself. As railroads required unprecedented capital in the mid-19th century, the limited liability corporation emerged to make ownership divisible \cite{Fordham2018CorporateForm, HBS2010RailroadFinance}. Companies—with distinct management, assets, and competitive positions—were transformed into standardized shares through listing requirements, accounting standards, and metrics like P/E ratios.

This pattern repeats across commodities. Grain markets suffered chronic quality disputes until USDA grading and futures contracts imposed standardization \cite{USDAERS2025}. Oil moved from chaotic local trade to global benchmarks through standards and reference prices \cite{stern2020comparative}. Table 1 traces this pattern across asset classes.

\small
\setlength{\tabcolsep}{4pt}
\rowcolors{2}{lightgray}{white}
\begin{longtable}{%
>{\raggedright\arraybackslash}p{0.16\textwidth}
>{\raggedright\arraybackslash}p{0.26\textwidth}
>{\raggedright\arraybackslash}p{0.20\textwidth}
>{\raggedright\arraybackslash}p{0.12\textwidth}
>{\raggedright\arraybackslash}p{0.24\textwidth}}
\caption{Historical asset market development: From heterogeneity to standardization}
\label{tab:historical-assets}\\
\toprule
\textbf{Asset Class} & \textbf{Mechanism} & \textbf{Developer} & \textbf{Date(s)} & \textbf{Outcome} \\
\midrule
\endfirsthead
\toprule
\textbf{Asset Class} & \textbf{Mechanism} & \textbf{Developer} & \textbf{Date(s)} & \textbf{Outcome} \\
\midrule
\endhead
\bottomrule
\endfoot
Corporate Equity & Limited liability corporations and standardized shares: divisible ownership with formalized listing requirements &
Railroad companies; New York Stock Exchange & 1850s--1860s &
Enabled capital mobilization and liquid markets for company ownership \\
Agriculture & Grain futures and warehouse receipts: standardized grading and storage contracts &
Chicago Board of Trade (CBOT) & 1848 &
Stabilized food supply chains and enabled hedging against harvest volatility \\
Oil & Futures contracts and spot benchmarks: standardized delivery and pricing &
Joseph Leiter; Chicago Board of Trade (CBOT); later NYMEX, OPEC & 1870s--1900s &
Shifted oil from local commodity to globally benchmarked and traded resource \\
Data & Emerging proposals for standardized data valuation, licensing, and registries &
Being established & 2020s--Present &
Transforming data from byproduct to recognized capital asset \\
\end{longtable}
\normalsize

\subsection{The state of research on data economics}

\textbf{Technical foundations from AI research.} Machine learning research has established empirical scaling laws demonstrating predictable relationships between training data volume, compute resources, and model performance \cite{kaplan2020scaling}. Work on data quality and curation reveals that not all data contributes equally—heterogeneity creates variation in marginal value. However, this research measures contribution in technical metrics (perplexity, accuracy) rather than economic units (prices, marginal products), leaving a translation gap.

\textbf{Economic theory on nonrivalry and market structure.} Economics research has focused on data's distinctive properties as a nonrival good. Jones and Tonetti \cite{jones2020nonrivalry} show that nonrivalry generates increasing returns to scale and creates tension between private incentives and social efficiency. Farboodi and Veldkamp \cite{farboodi2021long} model data as endogenously valuable information. Analysis of market structure reveals how information asymmetries impede exchange: Bergemann and Bonatti \cite{bergemann2024datamarkets} show that uncertainty about data value creates adverse selection; Santesteban and Longpre \cite{Santesteban2020-ko} document how heterogeneity creates barriers to entry that concentrate market power.

\textbf{Legal and regulatory frameworks.} Privacy regulations like GDPR establish individual rights over personal data but without mechanisms for collective action or market exchange \cite{acquisti2016economics}. Legal scholarship has proposed alternatives: Delacroix and Lawrence \cite{delacroix2019bottom} advocate bottom-up data trusts for collective governance; Pentland \cite{pentland2020neweconomy} proposes treating data as tradable capital. While some implementations are emerging, scalable governance structures with robust enforcement mechanisms and standardized metrics remain underdeveloped.

\textbf{Persistent gaps.} Three challenges emerge: (1) no consensus on appropriate units for measuring and pricing data—AI research measures tokens and parameters; economics discusses datasets and streams; regulation addresses individual records; (2) incomplete understanding of how data quality, quantity, and combination affect value, particularly how datasets combine to create value; (3) production function frameworks do not adequately represent data as a distinct input or capture its complementarities with compute and labor. 

\section{Data as a composable unit of production}

No consensus exists on the appropriate units for measuring, pricing, or exchanging data. AI research measures tokens; economics discusses datasets; regulation addresses individual records. Table~\ref{tab:data-hierarchy} proposes a unified hierarchy of exchangeable units and maps how different pricing mechanisms emerge at each level. While not yet standardized in practice, this provides a conceptual foundation for understanding different pricing mechanisms and exchange arrangements.

\begin{table}[H]
\centering
\caption[Atomic hierarchy of data as exchangeable units.]{Atomic hierarchy of data
as exchangeable units.\tablefootnote{These units can also be aggregated
collectively, introducing another dimension of exchange where data from different
sources is pooled and transacted as a shared asset~\cite{pentland2020neweconomy}.}}
\label{tab:data-hierarchy}
\small
\rowcolors{2}{lightgray}{white}
\begin{tabularx}{\textwidth}{
>{\raggedright\arraybackslash}p{0.10\textwidth}
>{\raggedright\arraybackslash}p{0.25\textwidth}
>{\raggedright\arraybackslash}p{0.41\textwidth}
>{\raggedright\arraybackslash}p{0.15\textwidth}}
\toprule
\textbf{Level} & \textbf{Unit of Exchange} & \textbf{Description} & \textbf{Typical Market Form} \\
\midrule

\textbf{Token} &
Smallest processable data fragment (e.g., tokenized text or scalar input) &
Divisible and composable\tablefootnote{Token-level pricing is now being operationalized through infrastructure such as Stripe’s usage-based billing API, which meters and charges per-token consumption via LLM proxies including OpenRouter, Cloudflare, Vercel, and Helicone. See also Coyle (2024) for broader economic-measurement framing.}; traded implicitly (e.g., through inference pricing); value tied to marginal compute cost. &
API pricing. \\

\textbf{Record} &
Single observation or labeled example &
Atomic contribution to learning; verification costly, so exchanged in bulk or via labor markets. &
Labeling platforms. \\

\textbf{Dataset} &
Curated collection of records &
Main tradable unit; value depends on quality, format, and domain specificity. &
Licensing, benchmarks. \\

\textbf{Corpus} &
Aggregate of datasets &
Compositional scale good; returns to diversity and coverage. &
Pretraining corpora. \\

\textbf{Stream} &
Continuous, time-ordered feed &
Dynamic input monetized by flow or access rather than ownership. &
Telemetry, feeds. \\
\bottomrule
\end{tabularx}
\normalsize
\end{table}
\section{Data in exchange: Transacting and pricing mechanisms}
\label{sec:pricing}

Despite data's growing economic importance, the heterogeneity established in Section~\ref{sec:properties} fundamentally obstructs standardized pricing. Transactions remain predominantly bilateral and bespoke, negotiated case by case based on data type, buyer context, and evolving legal boundaries. Most data transactions remain private; the examples documented here come from public announcements, company filings, and media reports, providing an incomplete but representative picture of market activity. Table~\ref{tab:data-pricing-alignment} maps principal pricing mechanisms to the data units established in Table~\ref{tab:data-hierarchy}. Table~\ref{tab:data-deals} in the Appendix provides systematic documentation of AI training data deals from 2020-2025, including deal structure, compensation terms, and exclusivity provisions across modalities.

\begin{figure}[htbp]
    \centering
    \includegraphics[width=0.8\linewidth]{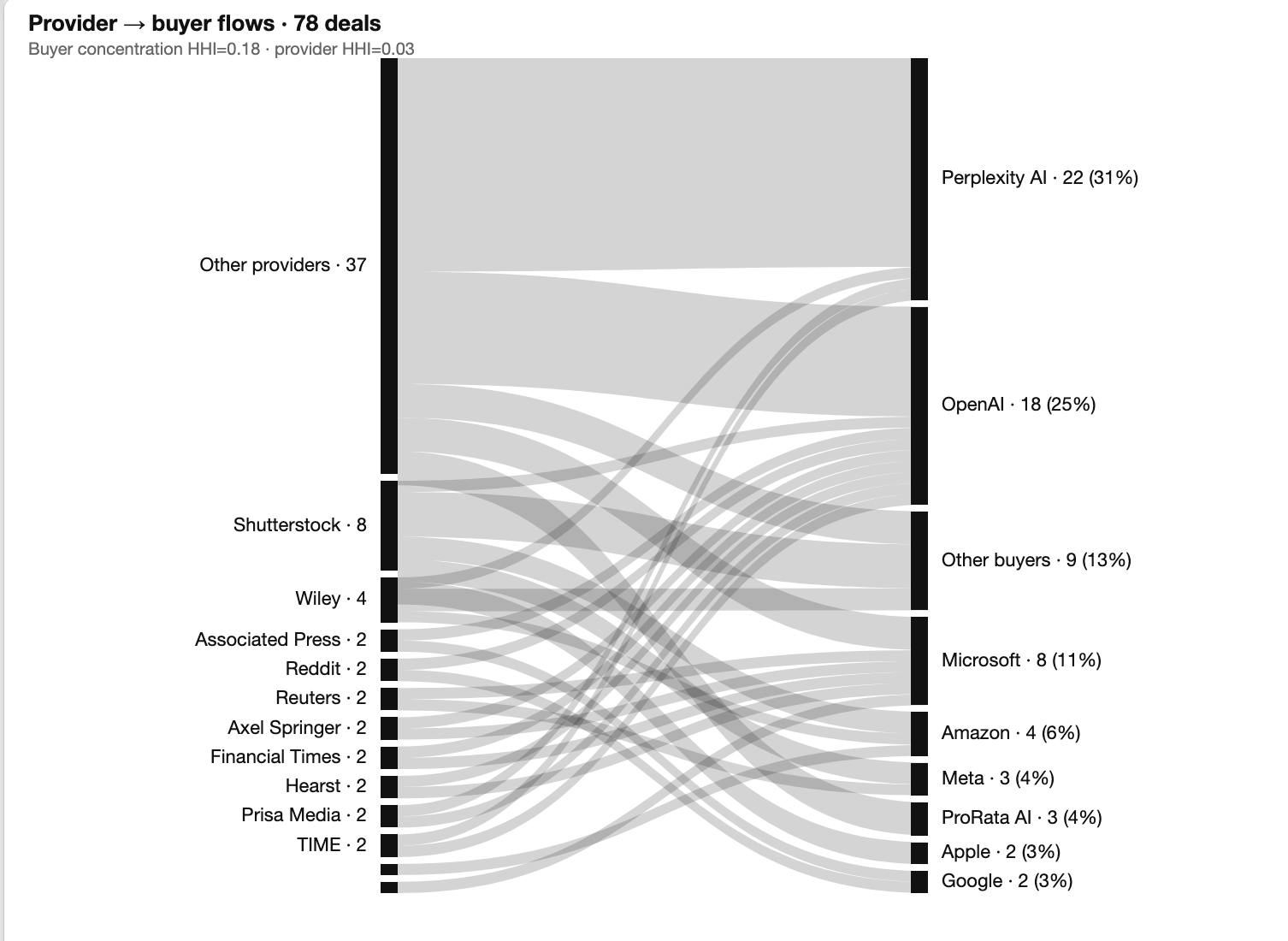}
    \caption{Sankey diagram visualizing provider-buyer flows across 78 AI training data deals, demonstrating a relatively concentrated market.}
    \label{fig:sankey}
\end{figure}

\small
\rowcolors{2}{lightgray}{white}
\begin{longtable}{
>{\raggedright\arraybackslash}p{0.18\textwidth}
>{\raggedright\arraybackslash}p{0.16\textwidth}
>{\raggedright\arraybackslash}p{0.38\textwidth}
>{\raggedright\arraybackslash}p{0.24\textwidth}}
\caption{Alignment between data units and pricing mechanisms}
\label{tab:data-pricing-alignment}\\
\toprule
\textbf{Pricing Mechanism} & \textbf{Applicable Unit(s)} & \textbf{Economic Logic} & \textbf{Examples} \\
\midrule
\endfirsthead
\toprule
\textbf{Pricing Mechanism} & \textbf{Applicable Unit(s)} & \textbf{Economic Logic} & \textbf{Examples} \\
\midrule
\endhead
\bottomrule
\endfoot

\textbf{Per-unit pricing} &
Token, Record &
Price proportional to usage volume or access frequency; reflects marginal processing or annotation cost. &
HarperCollins books (\$5k/title), indie music tracks (€0.30--€2/track), video footage (\$1--4/min) \\

\textbf{Aggregate licensing} &
Dataset, Corpus, Stream &
Payment for time-limited access rights to curated or proprietary data; includes both one-time and recurring subscriptions. Ownership remains with licensor. &
News Corp–OpenAI (\$250M+ over 5 years), Reddit–Google (\$60M/year), Dotdash Meredith–OpenAI (\$16M/year) \\

\textbf{Service-based pricing} &
Record, Dataset &
Payment for data transformation processes such as annotation, cleaning, or validation applied to existing data. &
Scale AI annotation platforms, data aggregation services \\

\textbf{Commissioning} &
Dataset, Corpus &
Upfront funding for new data collection or creation tailored to buyer specifications; pays for production process rather than existing assets. &
Mercor expert-generated domain data (\$450M ARR), custom video collection for vision models \\

\textbf{Open commons} &
All units &
Public datasets funded through government, research mandates, or voluntary contribution; provide competitive baseline for commercial markets. &
LAION/Common Crawl, Protein Data Bank \\

\end{longtable}
\normalsize

\textbf{Per-unit pricing} charges per discrete unit: licensing books at US\$5,000 each with 50/50 author splits, music at €0.30--€2.00 per track, and videos at US\$1--4 per minute. This mechanism prevails when units are clearly delineated, value per unit is consistent, and intermediaries (publishers, platforms) can aggregate creator content and negotiate on their behalf.

\textbf{Aggregate licensing} dominates large-scale enterprise deals: buyers pay fixed fees for time-limited access to curated corpora or feeds. Most licenses are non-exclusive---providers retain ownership and monetize simultaneous access to multiple buyers, but exclusivity like the News Corp deal commands price premiums. Deals frequently include hybrid components: News Corp receives both cash and substantial OpenAI API credits. These hybrid payments create lock-in: as providers integrate APIs into production systems, switching costs rise and relationships entrench.

\textbf{Service-based pricing} bundles data with transformation labor: platforms like Scale AI compensate crowd workers for annotation and curation, transforming raw data into labeled training sets. Buyers pay platforms for curated, quality-verified datasets. Platforms take a fee for coordination, quality assurance, and liability, then pay individual annotators.

\textbf{Commissioning} pays for \emph{new} data creation when required corpora don't exist. Mercor (US\$450 million ARR) connects AI labs with domain experts to generate specialized training data \cite{zeff2025mercor}; independent video creators supply custom footage to AI video labs. Pricing typically follows a consulting model: hourly rate for expert labor plus upcharge for the platform providing coordination, curation and quality assurance. 

\textbf{Open commons as competitive baseline.} Public datasets—LAION/Common Crawl for text and images, ESA's Copernicus satellite data, the Protein Data Bank for structural biology, and publicly funded research data from institutions like Germany's DLR—provide a competitive floor for commercial data markets. These commons exist due to public funding, regulatory mandates, or voluntary contribution, and serve as substitutes that discipline pricing in private markets. Recognition of data as a national strategic asset has accelerated investment in high-quality scientific datasets \cite{whitehouse2025americanaiactionplan}, further expanding the commons baseline.

\textbf{Implicit data exchanges} constitute a fifth, often-overlooked category. Here, platforms provide free or subsidized services---language models trained on conversations, recommendation systems trained on user behavior---in exchange for data rights outlined in the terms of service. Critically, this differs from advertising: users are not the product sold to advertisers; rather, their data inputs are harvested to train AI models sold to third parties. Reddit's retrospective licensing of user-generated content to Google (US\$60 million annually) exemplifies the pattern: platforms accumulate vast corpora through terms-of-service data grants, then monetize that corpus. 

Data deals generally exclude most data creators: of 24 major deals in Table~\ref{tab:data-deals}, only 7 compensate the original creator of the data, while the remaining 17 (News Corp's journalists, Reddit's users, Wiley's researchers) reward only intermediaries. This stems from scale barriers---individual creators lack negotiating power, falling below labs' minimum thresholds. Platforms aggregate user content and capture revenues. Le Monde's 25\% journalist revenue share, achieved through union leverage, remains a notable exception. Emerging data unions are experimenting with collective bargaining to address this, though implementation remains nascent. Per-unit pricing and commissioning do allow direct creator payments---HarperCollins' 50/50 author splits, indie music at €0.30--€2 per track, Mercor's expert hourly rates.

\section{Representing data in the production function}

Having established data's properties, its role in exchange, and the pricing mechanisms that govern transactions, we now address how data should be represented in economic production functions. Current growth and production models do not explicitly include data as a distinct factor, yet AI production depends heavily on it. This section outlines why explicit representation matters and what fundamental unknowns remain unresolved.

\subsection{The framework}

Current production function formulations do not explicitly model data as a distinct input. Rather than proposing a specific functional form, we argue that data should be treated as a distinct input. We represent this conceptually as:

\begin{equation}
Y = f(K, L, D, A)
\end{equation}

where $Y$ represents output, $K$ is capital (including compute infrastructure), $L$ is labor, $D$ is data, and $A$ captures technology and algorithmic efficiency. This notation indicates that data should be treated as a distinct factor of production rather than subsumed under capital, labor, or technology—but does not commit to a specific functional form, parametrization, or substitution elasticities.

Data could theoretically be incorporated into existing terms: as capital $K$ (an acquired asset), labor $L$ (effort in collection and processing), or technology $A$ (information improving productivity). However, doing so obscures data's distinctive properties: nonrivalry, context-dependence, and emergent rivalry through contamination and staleness. Data behaves fundamentally differently from traditional inputs. 

Unlike capital, it does not depreciate through use but may lose value through replication or obsolescence. Unlike labor, it can be used simultaneously by multiple producers without depletion. Unlike pure technology, data must be acquired—often at significant cost—and exhibits heterogeneity such that composition and complementarities affect its marginal contribution. Explicit representation enables analysis of data markets, optimal investment, and measurement of data's contribution to productivity growth. We remain agnostic about whether data follows Cobb-Douglas, CES, or other functional forms—that is an empirical question subsequent research should address. Our contribution here is establishing that data should appear explicitly in production function representations rather than being absorbed into other terms.

\subsection{Data's contribution depends on how it combines with other factors.} We observe both complementarity and substitutability. Frontiers in AI demonstrate strong complementarity: training state-of-the-art models requires both massive amounts of data and unprecedented compute. To some extent, firms can substitute compute for data through synthetic data generation, although it cannot fully replace diverse, real-world data for capturing edge cases and novel domains. The elasticities of these relationships, and how they vary by domain and task, remain empirically undetermined, but trade offs exist in practice in industry. 

Returns to scale on data represent another critical dimension. Empirical computer science research on language model training demonstrates that performance improves as a power law in dataset size, with exponents typically between 0.3 and 0.5 \cite{kaplan2020scaling, hoffmann2022training}, suggesting diminishing marginal returns from a model performance perspective. However, these estimates rest on limited data regimes, as researchers face a ``data wall'' around 15 trillion tokens of public internet text \cite{villalobos2024data}, beyond which empirical evidence becomes sparse. We do not know whether diminishing returns hold beyond this boundary, whether specialized or high-quality data exhibits different characteristics, or whether technical scaling laws map to economic returns once quality effects, network dynamics, and contamination risks are factored in.

\begin{figure}[H]
\centering
\begin{tikzpicture}
\begin{axis}[
    width=0.82\linewidth,
    height=0.52\linewidth,
    xlabel={Data Volume/Usage},
    ylabel={Marginal Value},
    xmin=0, xmax=10.4,
    ymin=0, ymax=26,
    xtick={0,2,4,6,8,10},
    ytick={0,5,10,15,20,25},
    axis lines=left,
    grid=major,
    grid style={draw=black!15, line width=0.3pt},
    tick align=outside,
    tick style={black!50},
    label style={font=\small},
    tick label style={font=\small},
    legend style={
        at={(0.03,0.97)}, anchor=north west,
        draw=none, fill=none, font=\small, row sep=1pt,
    },
    legend cell align=left,
    domain=0.05:10,
    samples=200,
    every axis plot/.append style={line width=1pt},
]
\addplot[black, solid] {10*ln(1+x)};
\addlegendentry{Capital-like (diminishing returns)}

\addplot[black!55, dash pattern=on 6pt off 3pt] {3*sqrt(x)};
\addlegendentry{Knowledge-like (increasing returns)}

\addplot[black!55, dash pattern=on 1pt off 2.5pt] {15*exp(-(x-5)^2/10)};
\addlegendentry{Contaminated/overuse (inverted-U)}
\end{axis}
\end{tikzpicture}
\caption{Three stylized models for data's contribution to AI production: diminishing returns (capital-like), sustained or increasing returns with quality, and inverted-U under contamination or overuse.}
\label{fig:returns}
\end{figure}

\textbf{Data's role differs across the machine learning pipeline.} Pre-training predominantly uses large-scale publicly available datasets, where volume drives value. Post-training prioritizes high-quality curated datasets commanding premium prices. Inference generates continuous user feedback. This creates segmented markets with distinct pricing dynamics, but whether these reflect fundamental features of AI production or contingencies of current technology remains unclear.

\begin{figure}[H]
\centering
\includegraphics[width=0.9\textwidth]{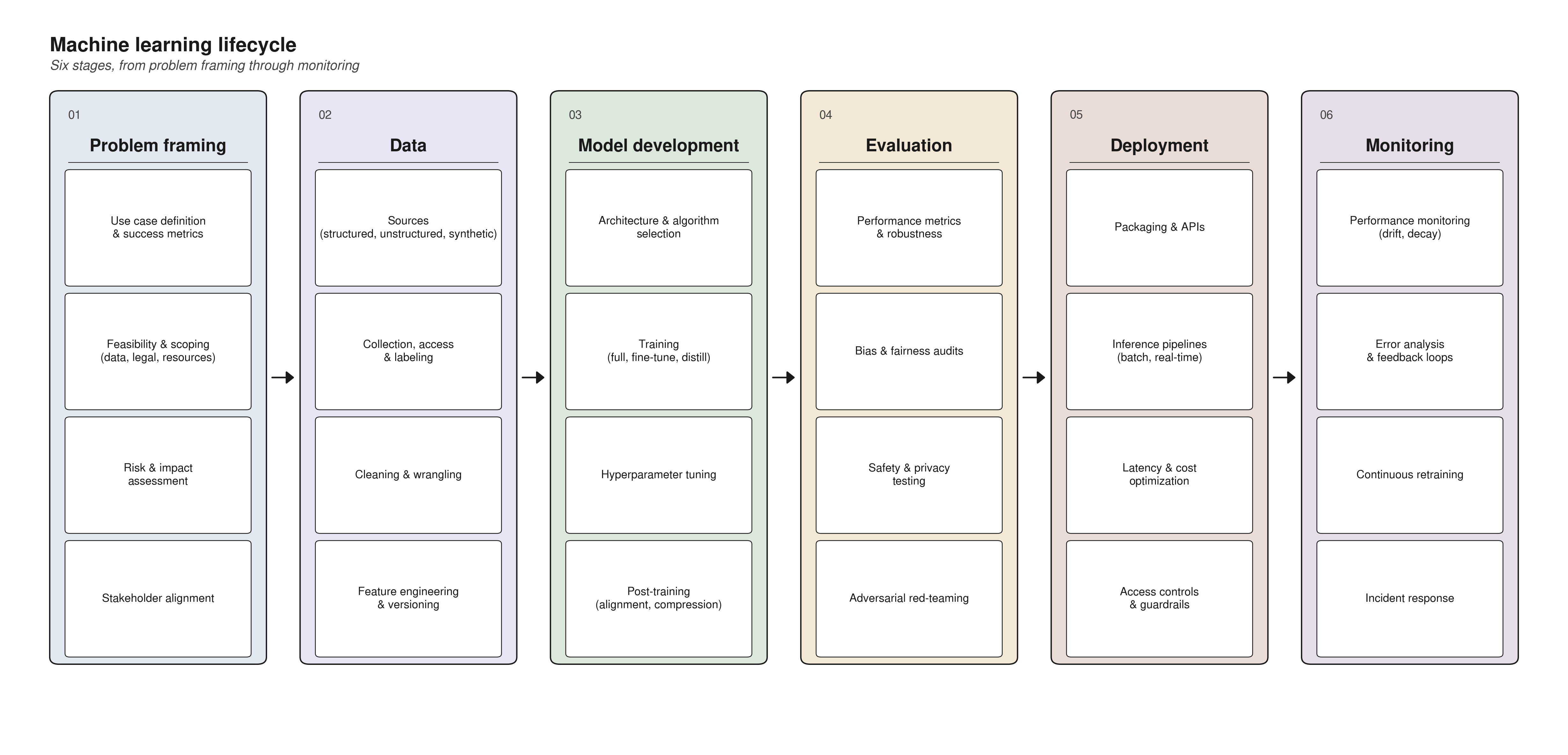}
\caption{Visual of the machine learning pipeline.}
\label{fig:ml-lifecycle}
\end{figure}

\section{Open Problems in Data Economics}
\label{sec:open-problems}

This paper establishes a framework for understanding data as a distinct factor of production with unique economic properties: nonrivalry, context-dependence, composability, and emergent rivalry through contamination. We have shown how data can be decomposed into exchangeable units, documented the pricing mechanisms that govern current transactions, and argued for explicit representation in production functions. However, building a complete theory of data economics requires sustained empirical and theoretical work across multiple disciplines.

We conclude by posing foundational questions that emerge from this framework. These questions are interdisciplinary by nature---addressing them requires collaboration among economists, computer scientists, legal scholars, and policymakers. Progress on these questions will determine whether data markets can function efficiently and whether we can properly account for data's contribution to economic growth.

\textbf{1. How do we measure and value data when its worth depends on context, composition, and who has access?}

Data's value is not intrinsic but depends on what other data exists, how it will be used, and who can access it. The same dataset has different value for different buyers with different existing corpora and different applications, and depends on which other buyers have already accessed the dataset. Exclusive access commands premium pricing; value degrades as access spreads. Technical metrics (tokens, records) do not map to economic value; quality is task-specific and compositional. Addressing this requires developing measurement frameworks that capture context-dependence, valuation models under interdependent preferences, and mechanisms for price discovery when inspection enables copying. 

\textbf{2. What property rights and governance mechanisms support efficient data allocation while preserving privacy and preventing monopolization?}

Individual ownership (GDPR-style) protects privacy but makes efficient pooling challenging; platform ownership enables scale but concentrates control; open commons maximize access but reduce incentives for data creators to contribute and may limit sustainability. Data's nonrivalry means multiple parties can use the same data simultaneously, but excludability is essential for market formation. Addressing this requires legal frameworks for data rights (law), institutional designs for data trusts and cooperatives (mechanism design), technical infrastructure for privacy-preserving computation and provenance tracking (computer science), and welfare analysis of alternative governance regimes (policy/economics). Historical precedents show that property rights choices had lasting effects on market structure for oil, spectrum, and other assets; the choices we make for data will similarly shape long-run outcomes.

\textbf{3. How do we empirically estimate data's contribution to production and productivity?}

We have argued that data should appear explicitly in production functions rather than being absorbed into capital, labor, or technology. But we have not specified what functional form data takes, how it enters production, or what its mathematical properties are. Building this theory requires specifying whether data follows Cobb-Douglas, CES, or other functional forms; deriving the marginal product of data and its elasticities with other inputs; and working through the implications for substitution, complementarity, and returns to scale. An ideal starting point would be empirical data on what training datasets each AI firm uses, their model performance, and revenue outcomes—this would enable both theory development and direct estimation.

Such firm-level data does not exist publicly. Progress instead requires combining controlled experiments isolating data's causal effects on model performance, natural experiments where data access varies, and evidence on firms' data investment decisions. These can answer concrete questions: How does model performance scale with dataset size and composition? Where do diminishing returns emerge? Can compute and labor substitute for data, or is data complementary and constraining? Do relationships differ across pre-training versus fine-tuning? Once empirical patterns emerge, they can guide theoretical specification and lead toward a functioning theory of data as a production input.

\textbf{4. Can we design markets and mechanisms for heterogeneous, compositional goods?}

Data's value depends on buyers' existing holdings and how datasets combine in training—not on the dataset in isolation. Buyers cannot assess value without examining data, but examination makes copying possible. Attribution is computationally intractable when models train on millions of sources. These challenges require developing new market mechanisms that handle interdependent valuations and compositional effects (mechanism design), standards for provenance and quality certification that enable price discovery without full inspection (institutional design), attribution methods that are theoretically sound and computationally feasible (computer science and economics), and minimal infrastructure for market formation without centralizing control (policy and industry coordination). Historical precedents show that market formation for grain, oil, and equities required developing measurement standards and exchange institutions; data markets need analogous infrastructure.


\section{Conclusion}

Data has become the decisive input shaping AI production, competition, and growth, yet it remains the least formalized in economic analysis. This paper outlines the foundations for a field of data economics: representing data as a distinct factor of production, documenting how it is priced and exchanged, and defining the open problems that must be solved for efficient markets to emerge.

As data takes its place alongside labor and capital as a primary scarce input, the stakes are not just about optimizing AI production but about deciding who owns the means of intelligence itself. Economists must formalize production functions that capture data’s nonrival and compositional nature; computer scientists must quantify its contribution to performance and develop verifiable provenance systems; legal and policy scholars must construct governance frameworks that balance privacy, innovation, and accessibility to keep the ecosystem healthy. As none of these tasks can succeed in isolation, a shared nomenclature must be built to lay the foundations for the shared infrastructure that has yet been formed. 

\bibliography{main}

@article{Hopkins_2025,
   title={AI Supply Chains: An Emerging Ecosystem of AI Actors, Products, and Services},
   volume={8},
   ISSN={3065-8365},
   url={http://dx.doi.org/10.1609/aies.v8i2.36628},
   DOI={10.1609/aies.v8i2.36628},
   number={2},
   journal={Proceedings of the AAAI/ACM Conference on AI, Ethics, and Society},
   publisher={Association for the Advancement of Artificial Intelligence (AAAI)},
   author={Hopkins, Aspen and Cen, Sarah H. and Struckman, Isabella and Ilyas, Andrew and Videgaray, Luis and Mądry, Aleksander},
   year={2025},
   month=Oct, pages={1266–1277} }

@article{hoffmann2022training,
  title={Training compute-optimal large language models},
  author={Hoffmann, Jordan and Borgeaud, Sebastian and Mensch, Arthur and Buchatskaya, Elena and Cai, Trevor and Rutherford, Eliza and Casas, Diego de Las and Hendricks, Lisa Anne and Welbl, Johannes and Clark, Aidan and others},
  journal={arXiv preprint arXiv:2203.15556},
  year={2022}
}

@misc{anthropic2024economic,
  author       = {{Anthropic}},
  title        = {Economic Index Geography},
  year         = {2024},
  howpublished = {\url{https://www.anthropic.com/research/economic-index-geography}},
  note         = {Accessed: 2025-10-07}
}

@misc{anthropic2024futures,
  author       = {{Anthropic}},
  title        = {Economic Futures Project},
  year         = {2024},
  howpublished = {\url{https://www.anthropic.com/economic-futures}},
  note         = {Accessed: 2025-10-07}
}

@misc{stripe2024fellowship,
  author       = {{Stripe}},
  title        = {Economics of AI Fellowship},
  year         = {2024},
  howpublished = {\url{https://stripe.events/fellowship}},
  note         = {Accessed: 2025-10-07}
}

@misc{openai2024gdpval,
  author       = {{OpenAI}},
  title        = {GDPval: Evaluating Model Capabilities on Economically Valuable Tasks},
  year         = {2024},
  howpublished = {\url{https://openai.com/index/gdpval/}},
  note         = {Accessed: 2025-10-07}
}

@article{farboodi2021long,
  title={Long-run growth of financial data technology},
  author={Farboodi, Maryam and Veldkamp, Laura},
  journal={American Economic Review},
  volume={111},
  number={8},
  pages={2485--2523},
  year={2021}
}

@misc{villalobos2024data,
  author       = {Villalobos, Pablo and Ho, Anson and Sevilla, Jaime and Besiroglu, Tamay and Heim, Lennart and Hobbhahn, Marius},
  title        = {Will we run out of data? Limits of LLM scaling based on human-generated data},
  year         = {2024},
  howpublished = {\url{https://epoch.ai/blog/will-we-run-out-of-data-limits-of-llm-scaling-based-on-human-generated-data}},
  note         = {Accessed: 2025-10-07}
}

@misc{openminedAIdata,
  author       = {{OpenMined}},
  title        = {AI is trained and evaluated on less than 0.01\% of the world's data},
  year         = {2025},
  howpublished = {\url{https://openmined.org/}},
  note         = {Accessed: 2025-10-07}
}

@misc{wikipediaDarkData,
  author       = {{Wikipedia contributors}},
  title        = {Dark data},
  year         = {2025},
  howpublished = {\url{https://en.wikipedia.org/wiki/Dark_data}},
  note         = {Accessed: 2025-10-07}
}

@ARTICLE{Santesteban2020-ko,
  title     = "How big data confers market power to big tech: Leveraging the
               perspective of data science",
  author    = "Santesteban, Cristian and Longpre, Shayne",
  journal   = "Antitrust Bull.",
  publisher = "SAGE Publications",
  volume    =  65,
  number    =  3,
  pages     = "459--485",
  month     =  sep,
  year      =  2020,
  language  = "en"
}

@misc{reuel2025openproblemstechnicalai,
      title={Open Problems in Technical AI Governance}, 
      author={Anka Reuel and Ben Bucknall and Stephen Casper and Tim Fist and Lisa Soder and Onni Aarne and Lewis Hammond and Lujain Ibrahim and Alan Chan and Peter Wills and Markus Anderljung and Ben Garfinkel and Lennart Heim and Andrew Trask and Gabriel Mukobi and Rylan Schaeffer and Mauricio Baker and Sara Hooker and Irene Solaiman and Alexandra Sasha Luccioni and Nitarshan Rajkumar and Nicolas Moës and Jeffrey Ladish and David Bau and Paul Bricman and Neel Guha and Jessica Newman and Yoshua Bengio and Tobin South and Alex Pentland and Sanmi Koyejo and Mykel J. Kochenderfer and Robert Trager},
      year={2025},
      eprint={2407.14981},
      archivePrefix={arXiv},
      primaryClass={cs.CY},
      url={https://arxiv.org/abs/2407.14981}, 
}

@misc{choi2025contamination,
  title        = {How Contaminated Is Your Benchmark? Quantifying and Mitigating Data Leakage in LLM Evaluation},
  author       = {Choi, Minseok and Lee, Sangwoo and Kim, Taewook and Park, Joonseok and Kim, Minjoon and Kim, Heewon and Lee, Kyungjae and Kim, Dongha},
  year         = {2025},
  eprint       = {2502.00678},
  archivePrefix= {arXiv},
  primaryClass = {cs.CL},
  url          = {https://arxiv.org/abs/2502.00678},
  note         = {Proposes systematic methods to detect benchmark leakage and quantify contamination in evaluation datasets.}
}

@misc{huang2023poisoningsurvey,
  title        = {Data Poisoning in Deep Learning: A Survey},
  author       = {Huang, Siyuan and Zhu, Yihan and Wang, Zekun and Xue, Ming and He, Zhenhua},
  year         = {2023},
  eprint       = {2302.10149},
  archivePrefix= {arXiv},
  primaryClass = {cs.LG},
  url          = {https://arxiv.org/abs/2302.10149},
  note         = {Comprehensive survey of adversarial data poisoning attacks and defenses in deep learning.}
}

@misc{li2025preferenceleakage,
  title        = {Preference Leakage: A Contamination Problem in LLM-as-a-Judge},
  author       = {Li, Xiang and Wang, Tianhao and Chen, Kai and Wu, Yikai and Zhao, Jiaming and Zhu, Yuchen},
  year         = {2025},
  eprint       = {2502.01534},
  archivePrefix= {arXiv},
  primaryClass = {cs.CL},
  url          = {https://arxiv.org/abs/2502.01534},
  note         = {Identifies feedback loop contamination in human preference data for LLM-as-a-judge pipelines.}
}

@misc{longpre2024consent,
  title        = {Consent in the Age of Large Language Models: The Case for Data Dignity},
  author       = {Longpre, Shayne and Stiennon, Nisan and Raffel, Colin and Hooker, Sara and Jernite, Yacine},
  year         = {2024},
  eprint       = {2410.11294},
  archivePrefix= {arXiv},
  primaryClass = {cs.CY},
  url          = {https://arxiv.org/abs/2410.11294},
  note         = {Analyzes consent-based data access protocols and their implications for AI training data supply.}
}

@misc{zhang2024carefulexaminationlargelanguage,
  title        = {A Careful Examination of Large Language Model Performance on Grade School Arithmetic},
  author       = {Zhang, Hugh and Da, Jeff and Lee, Dean and Robinson, Vaughn and Wu, Catherine and Song, Will and Zhao, Tiffany and Raja, Pranav and Zhuang, Charlotte and Slack, Dylan and Lyu, Qin and Hendryx, Sean and Kaplan, Russell and Lunati, Michele and Yue, Summer},
  year         = {2024},
  eprint       = {2405.00332},
  archivePrefix= {arXiv},
  primaryClass = {cs.CL},
  url          = {https://arxiv.org/abs/2405.00332},
  note         = {Examines benchmark contamination and memorization in arithmetic reasoning tasks.}
}

@ARTICLE{Huang2021-or,
  title     = "Toward a research framework to conceptualize data as a factor of
               production: The data marketplace perspective",
  author    = "Huang, Lihua and Dou, Yifan and Liu, Yezheng and Wang, Jinzhao
               and Chen, Gang and Zhang, Xiaoyang and Wang, Runyin",
  journal   = "Fundam. Res.",
  publisher = "Elsevier BV",
  volume    =  1,
  number    =  5,
  pages     = "586--594",
  year      =  {2021},
  copyright = "http://creativecommons.org/licenses/by-nc-nd/4.0/",
  language  = "en"
}

@book{pentland2020neweconomy,
  author    = {Pentland, Alex},
  title     = {Building the New Economy: Data as Capital},
  publisher = {MIT Press},
  year      = {2020},
  address   = {Cambridge, MA},
  url       = {https://mitpress.mit.edu/9780262539736/building-the-new-economy/}
}

@article{jones2020nonrivalry,
  author  = {Jones, Charles I. and Tonetti, Christopher},
  title   = {Nonrivalry and the Economics of Data},
  journal = {American Economic Review},
  volume  = {110},
  number  = {9},
  pages   = {2819--2858},
  year    = {2020},
  doi     = {10.1257/aer.20191330}
}

@article{bergemann2024datamarkets,
  author  = {Bergemann, Dirk and Bonatti, Alessandro},
  title   = {Data Markets and the Economics of Information Flow},
  journal = {Annual Review of Economics},
  year    = {2024},
  volume  = {16},
  pages   = {129--155}
}

@article{delacroix2019bottom,
  author    = {Delacroix, Sylvie and Lawrence, Neil D.},
  title     = {Bottom-up data Trusts: disturbing the 'one size fits all' approach to data governance},
  journal   = {International Data Privacy Law},
  year      = {2019},
  volume    = {9},
  number    = {4},
  pages     = {236--252},
  doi       = {10.1093/idpl/ipz014}
}

@misc{ropesgray2025unfairuse,
  author       = {Mathews, Manav and {Ropes \& Gray LLP}},
  title        = {A Tale of Three Cases: How Fair Use Is Playing Out in AI Copyright Lawsuits},
  year         = {2025},
  howpublished = {\url{https://www.ropesgray.com/en/insights/alerts/2025/07/a-tale-of-three-cases-how-fair-use-is-playing-out-in-ai-copyright-lawsuits}},
  note         = {Published July 6, 2025}
}

@article{akerlof1970market,
  author  = {Akerlof, George A.},
  title   = {The Market for "Lemons": Quality Uncertainty and the Market Mechanism},
  journal = {The Quarterly Journal of Economics},
  year    = {1970},
  volume  = {84},
  number  = {3},
  pages   = {488--500},
  doi     = {10.2307/1879431}
}

@inproceedings{cohen2016feature,
  author    = {Maxime Cohen and Ilan Lobel and Renato Paes Leme},
  title     = {Feature-based Dynamic Pricing},
  booktitle = {Proceedings of the 2016 ACM Conference on Economics and Computation (EC '16)},
  year      = {2016},
  pages     = {647--664},
  publisher = {ACM},
  url       = {https://research.google/pubs/feature-based-dynamic-pricing/}
}

@article{kaplan2020scaling,
  author    = {Kaplan, Jared and McCandlish, Sam and Henighan, Tom and Brown, Tom B. and Chess, Benjamin and Child, Rewon and Gray, Scott and Radford, Alec and Wu, Jeffrey and Amodei, Dario},
  title     = {Scaling Laws for Neural Language Models},
  journal   = {arXiv preprint arXiv:2001.08361},
  year      = {2020},
  url       = {https://arxiv.org/abs/2001.08361}
}

@article{acquisti2016economics,
  title={The economics of privacy},
  author={Acquisti, Alessandro and Taylor, Curtis and Wagman, Liad},
  journal={Journal of Economic Literature},
  volume={54},
  number={2},
  pages={442--492},
  year={2016}
}

@misc{cbinsights2024licensing,
  author       = {{CB Insights}},
  title        = {AI Content Licensing Deals},
  year         = {2024},
  howpublished = {\url{https://www.cbinsights.com/research/ai-content-licensing-deals/}},
  note         = {Accessed: 2025-10-27}
}

@techreport{whitehouse2025americanaiactionplan,
  title        = {America's AI Action Plan},
  author       = {{Office of Science and Technology Policy}},
  institution  = {The White House, Executive Office of the President},
  year         = {2025},
  month        = jul,
  note         = {Released under the directive "Winning the AI Race"},
  url          = {https://www.whitehouse.gov/wp-content/uploads/2025/07/Americas-AI-Action-Plan.pdf}
}

@article{zeff2025mercor,
  title        = {Sources: AI Training Startup Mercor Eyes \$10B+ Valuation on \$450M Run Rate},
  author       = {Zeff, Maxwell and TechCrunch Editorial Team},
  year         = {2025},
  month        = sep,
  journal      = {TechCrunch},
  url          = {https://techcrunch.com/2025/09/09/sources-ai-training-startup-mercor-eyes-10b-valuation-on-450m-run-rate/},
  note         = {Mercor, a startup connecting AI labs with domain experts for model training, was reportedly in talks to raise Series C funding at a \$10B valuation.}
}

@misc{USDAERS2025,
  title     = {State Agricultural Trade by Country of Origin and Destination},
  author    = {United States Department of Agriculture, Economic Research Service},
  year      = {2025},
  url       = {https://ers.usda.gov/publications/pub-details?pubid=45732},
  note      = {Accessed October 27, 2025}
}

@techreport{stern2020comparative,
  title        = {A Comparative History of Oil and Gas Markets and Prices: Is 2020 Just an Extreme Cyclical Event or an Acceleration of the Energy Transition?},
  author       = {Stern, Jonathan and Imsirovic, Adi},
  institution  = {Oxford Institute for Energy Studies},
  type         = {Energy Insight 68},
  year         = {2020},
  month        = apr,
  url          = {https://www.oxfordenergy.org/wpcms/wp-content/uploads/2020/04/Insight-68-A-Comparative-History-of-Oil-and-Gas-Markets-and-Prices.pdf}
}

@article{Fordham2018CorporateForm,
  author = {Author not specified in snippet},
  title = {A Brief History of the Corporate Form and Why it Matters},
  journal = {Fordham Journal of Corporate \& Financial Law},
  year = {2018},
  url = {https://news.law.fordham.edu/jcfl/2018/11/18/a-brief-history-of-the-corporate-form-and-why-it-matters/#_ednref25},
  note = {Accessed October 27}
}

@misc{HBS2010RailroadFinance,
  title = {New Levels of Capitalism: Finance - Railroads and the Transformation of Capitalism},
  author = {Harvard Business School, Baker Library Historical Collections},
  year = {2010},
  url = {https://www.library.hbs.edu/hc/railroads/finance.html},
  note = {Accessed October 27, 2025}
}
\clearpage
\section{Appendix}

\vspace{-0.5em}
\begin{center}
\tiny
\setlength{\tabcolsep}{2pt}
\renewcommand{\arraystretch}{1.1}
\definecolor{lightgray}{gray}{0.95}

\rowcolors{2}{lightgray}{white}
\begin{longtable}{
>{\centering\arraybackslash}p{0.018\textwidth} 
>{\raggedright\arraybackslash}p{0.062\textwidth} 
>{\raggedright\arraybackslash}p{0.045\textwidth} 
>{\raggedright\arraybackslash}p{0.062\textwidth} 
>{\raggedright\arraybackslash}p{0.054\textwidth} 
>{\raggedright\arraybackslash}p{0.116\textwidth} 
>{\raggedright\arraybackslash}p{0.090\textwidth} 
>{\centering\arraybackslash}p{0.062\textwidth} 
>{\centering\arraybackslash}p{0.062\textwidth} 
>{\raggedright\arraybackslash}p{0.090\textwidth} 
>{\raggedright\arraybackslash}p{0.116\textwidth} 
>{\raggedright\arraybackslash}p{0.116\textwidth} 
}
\caption{AI Content Licensing and Data Deals (2020–2025)}
\label{tab:data-deals}\\
\toprule
\textbf{\#} & \textbf{Modality} & \textbf{Date} & \textbf{Provider} & \textbf{Buyer} & \textbf{Data Type} & \textbf{Reported Terms} & \textbf{Compensated?} & \textbf{Exclusive?} & \textbf{Pricing Mech.} & \textbf{Notable Details} & \textbf{Source(s)} \\
\midrule
\endfirsthead

\multicolumn{12}{@{}l}{\footnotesize AI Content Licensing and Data Deals (2020–2025)} \\
\toprule
\textbf{\#} & \textbf{Modality} & \textbf{Date} & \textbf{Provider} & \textbf{Buyer} & \textbf{Data Type} & \textbf{Reported Terms} & \textbf{Compensated?} & \textbf{Exclusive?} & \textbf{Pricing Mech.} & \textbf{Notable Details} & \textbf{Source(s)} \\
\midrule
\endhead

\midrule
\multicolumn{12}{r}{\footnotesize\textit{Continued on next page...}} \\
\endfoot

\bottomrule
\endlastfoot


1 & Text & 2
024-05-22 & News Corp & OpenAI & News archive \& publisher content (WSJ, The Times, NY Post) & $>$US\$250M / 5 yrs & No & Yes & Access licensing & Largest journalism-AI deal; cash + credits \cite{cbinsights2024licensing} & \href{https://investors.newscorp.com/news-releases/news-release-details/news-corp-and-openai-sign-landmark-multi-year-global-partnership}{NewsCorp, WSJ} \\

2 & Text & 2
024-02-22 & Reddit & Google & Social-media UGC feed & $\approx$US\$60M / yr & No & No & Volume-based access & Recurring API access for search \& training \cite{cbinsights2024licensing} & \href{https://www.reuters.com/technology/reddit-ai-content-licensing-deal-with-google-sources-say-2024-02-22/}{Reuters, The Verge} \\

3 & Text & 2024-05 & Dotdash Meredith & OpenAI & Magazine \& digital-media archives & $\ge$US\$16M / yr (fixed) & No & No & Access licensing (base + variable) & Includes legacy magazine brands \cite{cbinsights2024licensing} & \href{https://www.axios.com/2024/05/22/openai-news-corp-content-licensing-deal}{Axios} \\

4 & Text & 2024-11 & HarperCollins & Microsoft & Non-fiction book titles (AI training rights) & US\$5K / title; 50/50 split & Yes & No & Per-unit licensing & Early per-book pricing benchmark; limits verbatim output & \href{https://authorsguild.org/news/harpercollins-ai-licensing-deal/}{Authors Guild} \\

5 & Text & 2023 & Taylor \& Francis & Microsoft & Academic journals / textbooks & $\approx$US\$10M & Unclear & No & Access licensing (restricted) & License for academic content \cite{cbinsights2024licensing} & \href{https://www.thebookseller.com/news/wiley-set-to-earn-44m-from-ai-rights-deals-confirms-no-opt-out-for-authors}{The Bookseller} \\

6 & Text (Scientific) & 2024 & Wiley & Anthropic, AWS, Perplexity & Scientific content + metadata & US\$23M (2025 Proxy filing) & No & No & Limited-term structured license & Controlled-environment use \cite{cbinsights2024licensing} & \href{https://s27.q4cdn.com/812717746/files/doc_downloads/2025/2025-Proxy-Statement.pdf}{Wiley Proxy} \\

7 & Image/Video & 2021--2024 & Shutterstock & Meta, OpenAI, Google, Apple & Stock images + video + metadata & US\$25\textendash30M per deal (multi-yr) & Partial & No & Hybrid per-unit + access & Used for multimodal training; includes royalty fund & \href{https://venturebeat.com/ai/apples-25-50-million-shutterstock-deal-highlights-fierce-competition-for-ai-training-data/}{VentureBeat} \\

8 & Image & 2024 & Freepik & Unnamed AI firms &
$\approx$200M stock images &
$\approx$US\$6M (2--4\textcent{}/image) &
No & No &
Per-unit micro-licensing &
Large-scale low-cost imagery for pre-training &
\href{https://www.reuters.com/technology/inside-big-techs-underground-race-buy-ai-training-data-2024-04-05/}{Reuters} \\

9 & Image/Text & 2020--2023 & LAION / Common Crawl & Open model builders & Image-caption datasets + web crawls & No cash value & N/A & No & Commons / open-source & Foundational datasets (LAION-5B, Falcon, etc.) & \href{https://laion.ai/blog/laion-5b/}{LAION} \\

10 & Text (Media) & 2025-05 & Le Monde & OpenAI, Perplexity & News content licensing deal & Undisclosed (25\% rev share to journalists) & Yes & No & Access licensing & First to share AI revenue directly with journalists & \href{https://www.lemonde.fr/en/media/article/2025/06/09/le-monde-group-s-2024-accounts-mark-a-landmark-year_6742164_22.html}{Le Monde} \\

11 & Audio & 2025-07 & SourceAudio & ElevenLabs, Music.AI & Pre-cleared songs for AI training & US\$10M (multi-year deal) & Yes & No & Access licensing & Access to millions of licensed tracks & \href{https://www.recordoftheday.com/news-and-press/sourceaudio-teams-up-with-elevenlabs-as-preferred-partner-for-ai-music-licensing}{Record of the Day} \\

12 & Audio & 2024 & UMG, Warner & AI music startups (Suno, Mubert) & Music catalog rights (audio + lyrics) & Undisclosed & Unclear & No & Access licensing (music/audio) & Pilot licensing framework for generative music & \href{https://www.musicbusinessworldwide.com/universal-warner-explore-ai-licensing-pacts-2024/}{MBW} \\

13 & Audio & 2024 & Audius + indie labels & EU generative-music firms & Independent tracks \& stems & $\text{€0.30}-\text{€2.00}$ / track & Yes & No & Per-unit micro-licensing & Emerging artist-level licensing model & \href{https://www.billboard.com/pro/audius-eu-ai-music-licensing-deals-2024/}{Billboard} \\

14 & Video & 2025-01 & YouTube creators & OpenAI, Meta & Unpublished creator videos & $\approx$US\$5M total (US\$1\textendash4 / min) & Yes & No & Per-unit licensing (video minutes) & AI labs buying unpublished creator content & \href{https://petapixel.com/2025/01/13/ai-companies-are-paying-content-creators-for-their-unpublished-videos/}{PetaPixel} \\

15 & Video & 2023--24 & Independent creators & Runway, Pika Labs & Professional/unpublished video footage & $\approx$US\$1\textendash4 / min (est.) & Yes & No & Per-unit licensing & Used for vision-language model training & \href{https://the-decoder.com/openai-and-google-are-buying-youtubers-unpublished-videos-for-up-to-4-per-minute/}{The Decoder} \\

16 & Satellite & 2020--24 & Planet Labs & Agriculture / gov AI firms & Daily Earth observation imagery & $\approx$US\$180M annual rev & No & No & Subscription access licensing & Used for climate, agriculture \& defense models & \href{https://investors.planet.com/financials/annual-reports/}{Planet Labs} \\

17 & Health/Biotech & 2024 & Tempus & Pharma \& AI firms & Anonymized patient/genomic data & US\$200M / 3 yrs & No & No & Access licensing & Training medical LLMs; 40\% YoY growth \cite{cbinsights2024licensing} & \href{https://www.cbinsights.com/company/tempus-labs}{CB Insights} \\

18 & Corporate & 2025-06 & Scale AI & Meta & Equity stake for data services & US\$14.3B for 49\% stake & Variable & Yes & Strategic acquisition & Largest single data-related AI deal; Scale valued \$29B & \href{https://www.cnbc.com/2025/06/14/meta-invests-14b-scale-ai.html}{CNBC, FT} \\

19 & Corporate & 2025-04 & Informatica & Salesforce & Cloud data integration platform & $\approx$US\$8B & N/A & Yes & Full acquisition & Boosts enterprise AI data pipeline capabilities & \href{https://techcrunch.com/2025/04/12/salesforce-to-acquire-informatica-for-8b/}{TechCrunch} \\

20 & Legal/Books & 2025-09-05 & Authors \& Publishers (class) & Anthropic & Books (unauthorized) & US\$1.5B settlement ($\approx$US\$3K/book) & Yes & N/A & Legal settlement & Class action requiring data deletion \& future limits & \href{https://www.reuters.com/technology/anthropic-settlement-authors-2025-09-06/}{Reuters, WIRED} \\

21 & Text & 2025-08 & CuriosityStream & AI partners & Factual/documentary video library & US\$20\textendash30M / yr & No & No & Access licensing (subscription/API feed) & $\approx$25\% of 2025 revenue & \href{https://app.quotemedia.com/data/downloadFiling?webmasterId=131387\&ref=319353889\&type=PDF\&symbol=CURI\&companyName=CuriosityStream+Inc.\&formType=10-Q\&dateFiled=2025-08-06}{SEC Filing} \\

22 & Text & 2025 & New York Times & Amazon & Editorial news (NYT Cooking, The Athletic) & US\$20\textendash25M / yr & No & No & Access licensing & Enables Alexa to use Times content for summaries \& training & \href{https://www.geekwire.com/2025/report-amazon-to-pay-at-least-20m-a-year-in-ai-content-deal-with-new-york-times/}{GeekWire} \\

23 & Text & 2025 H1 & Chegg & AI partners & Expert-written Q\&A pairs (homework help database) & US\$11M (H1 2025) & Unclear & No & Access licensing & New revenue stream; $<$5\% of library & \href{https://www.sec.gov/Archives/edgar/data/1364954/000136495425000049/a9901-financialresultsq120.htm}{SEC Filing} \\

24 & Commissioning & 2022--2025 & Mercor & Mult. AI Labs & Domain expert talent for AI model training (RLHF, fine-tuning) & $\approx$US\$450M ARR & Yes & No & Commissioning / Service-based & Rapid growth; "Switzerland" provider & \href{https://techcrunch.com/2025/09/09/sources-ai-training-startup-mercor-eyes-10b-valuation-on-450m-run-rate/}{TechCrunch} \\

\end{longtable}
\end{center}
\normalsize
\end{document}